\documentclass{article}

\usepackage{arxiv}

\usepackage[utf8]{inputenc} 
\usepackage[T1]{fontenc}    
\usepackage{hyperref}       
\usepackage{url}            
\usepackage{booktabs}       
\usepackage{amsfonts}       
\usepackage{nicefrac}       
\usepackage{microtype}      
\usepackage{lipsum}
\usepackage{graphicx}
\graphicspath{ {./images/} }
\usepackage{braket}
\usepackage{amsmath}
\title{Infrared rainbow trapping via optical Tamm modes in one-dimensional dielectric chirped photonic crystals}

\author{
 Shailja Sharma \\
 School of Physical Sciences\\ National Institute of Science Education and Research\\ HBNI, Jatni - 752050, Odisha, India\\
  \texttt{shailja.sharma@niser.ac.in} \\
   \And
 Abhishek Mondal \\
  School of Physical Sciences\\ National Institute of Science Education and Research\\ HBNI, Jatni - 752050, Odisha, India\\
  \texttt{abhishek.mondal@niser.ac.in} \\
  \And
 Ritwick Das* \\
  School of Physical Sciences\\ National Institute of Science Education and Research\\ HBNI, Jatni - 752050, Odisha, India\\
  \texttt{ritwick.das@niser.ac.in} \\
}

\begin{document}
\maketitle
\begin{abstract}
The phenomenon of trapping broad spectrum of light is known as `rainbow trapping' and achieved using all-dielectric, hybrid metallo-dielectric or all-metallic configurations. In the latter cases, unavoidable ohmic losses result in sub-picosecond trapped-mode lifetimes. For all practical purposes, novel strategies are required to be devised for trapping and subsequently, releasing broadband electromagnetic (em) field with lifetime > 1 ps. We present a rainbow trapping configuration using the excitation of multiple optical Tamm (OT) modes in an one-dimensional chirped photonic crystal (PC) designed for adiabatically coupling counter-propagating modes. In the geometry, the multiple $\pi$ phase jumps enable excitation of OT modes when a thin plasmon-active metal is placed adjacent to the terminating layer of chirped-PC (CPC). The strongly localized OT resonances are spatially separated in the chirped-PC geometry and their group-velocities reduce to as low as $0.17c$. The time-domain simulations elucidate localization takes place in the dielectric sections of CPC which manifest into lifetimes $\sim 3~ps$.    
\end{abstract}


\section{Introduction}
Rainbow trapping is a phenomenon of slowing down (in principle stopping) a broadband optical radiation in an one- or two-dimensional stratified photonic system \cite{Shen:11,PhysRevApplied.12.024043,Toshihiko}. The excitation of surface-plasmon (SP) modes in periodic photonic crystal (PC) provides a plausible mechanism for trapping electromagnetic ($em$) radiation. Consequently, the configurations deployed for rainbow trapping are prominently metallo-dielectric or all-metallic \cite{PhysRevLett.102.056801,PhysRevB.80.161106,Chen2011,Gan5169}. Although significant reduction in group velocity ($v_g\leq~0.1c$) is achieved in such configurations, the absorption losses ($\alpha$) in plasmon-active metals limit the trapped-mode lifetime ($T_r$) in the visible spectral band \cite{Shen:11,PhysRevB.80.161106}. The impact turns more significant in the infrared (IR) band ($\geq~1~\mu m$ wavelength) where the absorption losses (by metals) is further high. Therefore, the recurring challenge is to develop plausible strategies which offer appreciably low loss along with small $v_g$ over broad bandwidth in metallo-dielectric configurations. All-dielectric photonic crystals (PCs) have been proposed for trapping light in the visible and IR band but they exhibit maximum $T_r\sim~0.2-0.3~ps$ which limits a majority of applications \cite{Shen:11,SHEN20113801}. The graded-$Si$ based composite gratings with $SiO_2$ spacer-layers offer a reliable platform for slowing down wavelengths $\geq~2~\mu m$ in the IR band \cite{Ghaderian}. Recently, through graphene-incorporated graded-$Si$ gratings, rainbow trapping in $10-50~THz$ range is realized with slowdown factors $\sim~0.001c$ \cite{Ghaderian}. However, the trapped plasmon modes still exhibit a sub-picosecond lifetime essentially due to higher losses. All-dielectric photonic crystals (PCs) offer a low-loss flexible platform for light-trapping through excitation of optical surface states. Such states could be excited at the interface of two different topologically non-trivial PCs or at the interface of a topologically non-trivial PC and homogeneous medium \cite{PhysRevB.76.165415,PhysRevB.74.045128,Kavokin}. Optical Tamm (OT) states, which exist at the interface between a plasmon-active metal and an one-dimensional (1D) PC (1D-PC), could be classified in the second category  \cite{PhysRevB.76.165415}. Optimally-designed geometries support OT states which are tightly confined at the PC-metal interface and exhibit a parabolic dispersion curve. Importantly, their dispersion lie within the light cone given by $k_{\vert\vert} = \frac{\omega}{c}$, where $k_{\vert\vert}$ is the in-plane wavevector component and $\omega$ is the angular frequency of light. Consequently, the transverse-electric (TE) as well as transverse-magnetic (TM) polarization could be excited through free-space coupling at normal incidence \cite{Chen:14,PhysRevB.76.165415,Sasin}. Due to the possibility of free-space coupling to OT modes, the coupling efficiencies are very high ($\geq~90\%$) which allows efficient device realization \cite{Auguie}. In the last decade, OT states/modes have attracted a wide-range of applications which include devising of optical sensors, narrowband tunable filters, optical switches, harmonic generators, slow-light devices etc. \cite{Maigyte,Mitsuteru,samir,Afinogenov,Symonds}. The nature of OT resonance in an 1D-PC yield localization of $em$-field at one resonant frequency only and therefore, any light trapping mechanism using OT states would be narrow band \cite{Symonds}. Many processes such as rainbow trapping or nonlinear non-degenerate frequency conversion primarily require multiple non-degenerate OT states in the same PC.       
\begin{figure}[htbp]
\centering\includegraphics[width=\linewidth]{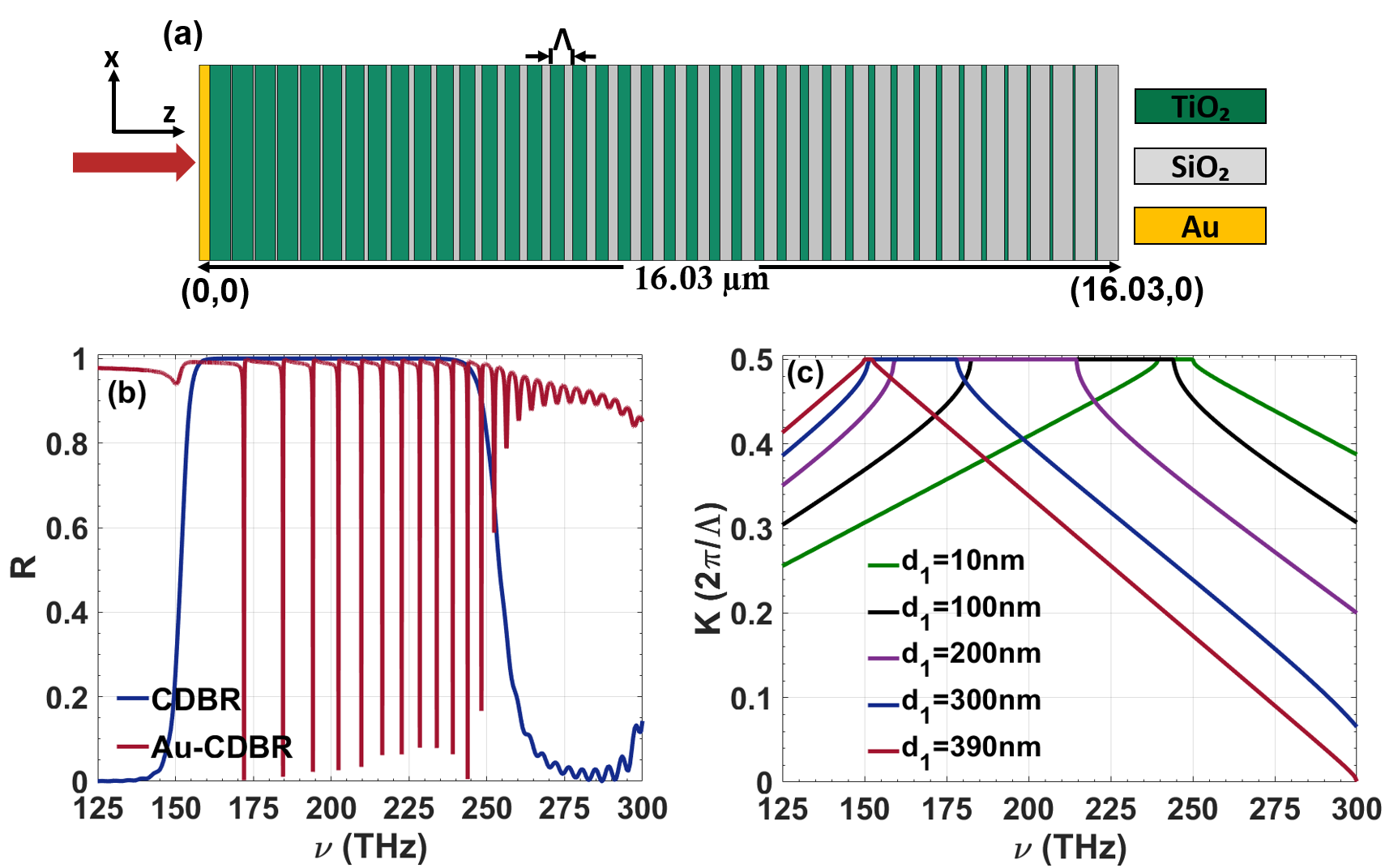}
\caption{a) A schematic of the chirped-PC geometry with a fixed periodicity ($\Lambda$) but variable duty cycle. b) Reflection spectrum of CPC (solid blue line) for $N=39$ unit cells and Au-CPC (red line). c) Represents dispersion relation \emph{i.e.} Bloch-wavevector ($K$) versus frequency ($\nu$) for periodic $TiO_2/SiO_2$ based PC geometries when $d_1 = 10~nm$ (green curve), $d_1 = 100~nm$ (black curve), $d_1 = 200~nm$ (purple curve), $d_1 = 300~nm$ (blue curve) and $d_1 = 390~nm$ (maroon line). All the PCs (with different $d_1$) have identical periodicity ($\Lambda = 400~nm$).}
\label{Figure1}
\end{figure}
Here, we present a 1D linearly chirped all-dielectric PC which support multiple optical Tamm-like modes. The OT modes are spatially separated within the PC and facilitate a broad spectrum to be trapped. The linearly-chirped PC has been designed such that the phase-mismatch ($\Delta \beta$) between the forward and backward propagating modes in the PC is `slowly' swept from a negative to a positive value along the propagation ($z$) direction, thus enabling adiabatic intermodal energy transfer \cite{yariv}. The chirped-PC (CPC) has a constant unit cell thickness ($\Lambda$) with monotonically varying duty-cycle \emph{i.e.} we consider two layers $A$ and $B$ of refractive indices $n_1$ and $n_2$ respectively. By considering $N$ unit cells in the PC, the thickness of layer $A$ in the $m^{th}$ unit cell is $d_{1m} = d_1- m\delta$ and that for layer $B$ is $d_{2m} = \Lambda - d_1 + m\delta$ (see Fig. \ref{Figure1}(a)). In order to adopt a perturbative approach to ascertain forward-backward coupling, we express the dielectric function for a periodic PC geometry as $\epsilon (x,y,z) = \epsilon (x,y) + \Delta \epsilon (x,y,z)$ where $\Delta \epsilon (x,y,z)$ is a periodic function of $z$. In presence of spatially-varying dielectric tensor, the coupled-mode equations are represented as \cite{yariv}.
\begin{equation} \label{eq:equation1}
  \frac{dA_b}{dz} = - i \frac{\beta_b}{|\beta_b|} {\kappa_s} A_f e^{i \Delta \beta z}  
\end{equation}
\begin{equation} \label{eq:equation2}
  \frac{dA_f}{dz} = - i \frac{\beta_f}{|\beta_f|} {\kappa_s^{*}} A_b e^{-i \Delta \beta z}  
\end{equation}
where $A_f$ ($\beta_f$) and $A_b$ ($\beta_b$) are the $z$-dependent complex amplitude (propagation constant) of forward propagating and backward propagating modes respectively and $\Delta \beta = 2 \beta \cos{\theta}-\frac{2m\pi}{\Lambda}$ defines the phase-mismatch. It is worth pointing out that $\beta = \beta_b = -\beta_f$ in a forward-backward mode-coupling scheme and $\theta$ is the angle of incidence. $\kappa_s$ defines the magnitude of coupling coefficient which utilizes the $s^{th}$ Fourier component of dielectric function for coupling the forward propagating mode to its backward propagating component. This is given by  
\begin{equation}\label{eq:equation3}
\kappa_s = \frac{\pi c}{2\lambda} \int \int {E_f^*}(x,y) \epsilon_s(x,y) {E}_b(x,y) dx dy
\end{equation}
where $\lambda$ is the operating wavelength, $\epsilon_s$ is the $s^{th}$ component of Fourier-series expansion of $\epsilon(x,y,z)$ mentioned above and $E_{f(b)}$ represent transverse mode-field distributions for forward (backward) propagating modes. Equations (\eqref{eq:equation1}) and (\eqref{eq:equation2}) could be transformed into a rotating-frame through the substitution \begin{equation}\label{eq:equation4}
    A_f = \Tilde{a_f} e^{-i/2[\Delta \beta(0) z - {\int_0}^z p(z') dz']}
\end{equation}
\begin{equation}\label{eq:equation5}
  A_b   = \Tilde{a_b} e^{i/2[\Delta \beta(0) z - {\int_0}^z p(z') dz']}
\end{equation}
where $\Delta \beta (0)$ is the phase-mismatch at $z=0$ and $p(z)$ is the modulation brought about by introducing chirp in the PC. This transformation yields $-i\frac{d}{dz}\ket{\Psi} = \hat{H} \ket{\Psi}$ where $\big| \Psi \big> = \bigg(\begin{matrix} \Tilde{a_f}\\
\Tilde{a_b}\end{matrix}\bigg)$ and $\hat{H} = \bigg(\begin{matrix} \Delta k & \kappa^{*}_s\\
-\kappa_s & -\Delta k \end{matrix}\bigg)$ where $\Delta k = \frac{\Delta \beta (0)-p(z)}{2}$. It is important to note that $\kappa_s^* = -\kappa_s$ and consequently, Hamiltonian $\hat{H}$ analogous to that representing a \emph{two-level atomic system} or a spin\emph{-1/2} particle in a homogeneous magnetic field \cite{eberly}. Subsequently, a state-vector $\rho~\equiv~[U,V,W]$ could be defined such that $U = \Tilde{a_f} {\Tilde{a_b}}^* + \Tilde{a_b}{\Tilde{a_f}}^*$, $V = -i[\Tilde{a_f} {\Tilde{a_b}}^* - \Tilde{a_b} {\Tilde{a_f}}^*]$ and $W = {| \Tilde{a_f}|}^2 - {| \Tilde{a_b}|}^2$. The state-vector $\rho$ could be mapped on an equivalent Bloch-sphere where the north-pole and the south-pole represents a purely forward-propagating and purely backward-propagating mode respectively \cite{eberly}. Any arbitrary point on the equivalent Bloch-sphere represents a state $\rho$ which is dictated by the values of $\Delta k$ and $\kappa$. For a given operating frequency ($\omega$), the coupling efficiency from a forward-propagating mode to a backward-propagating mode is given by $\eta=\frac{W+1}{2}$ and it maximizes when $\rho ~=~ [0,0,-1]$. This is conventionally achieved through enforcing $\Delta \beta = 0$ (\emph{Bragg's} condition) at the central photonic bandgap (PBG) frequency. In this equivalent `two-level' system, the propagation characteristics of a PC with well-defined periodicity could be described using the Bloch-wavevector ($K$) and it bears a relation
\begin{equation}
    K = \frac{m \pi}{\Lambda} \pm i \sqrt{\kappa^* \kappa - {\bigg(\frac{\Delta \beta }{2}\bigg)}^2}
\end{equation}
for $s = 1$ \cite{yariv}. Consequently, the width of PBG as well as the reflected beam amplitude is governed by the values of $\kappa$ and $\Delta k$. Since $K$ is complex within the PBG, then $|{\kappa}|^2 \geq (\frac{\Delta \beta}{2})^2$) defines the edges of PBG for the PC. In analogy with adiabatic population transfer in \emph{two-level} atomic system, a plausible route for maximizing $\eta$ is to vary $\Delta k$ (through $p(z)$) such that the adiabatic constraint given by $ \frac{d\Delta k}{dz} << {|\kappa|}^2$ is obeyed at each $z$. The longitudinal variation in $\Delta k$ is brought about by a linear chirp in thickness of layers $A$ and $B$ described before. The CPC configuration which ensures adiabatic following has two distinct features. (a) PBG for an optimally CPC increases (b) CPC exhibits \emph{multiple} phase-jumps  ([$0\rightarrow\pi$] or [$0\rightarrow -\pi$]) within the PBG. These points will be apparent in the example considered below. 

Let us consider layer $A\equiv TiO_2$ and layer $B\equiv SiO_2$. The linear chirp leads to a longitudinally varying \emph{average} refractive index ($\Bar{n} = \sqrt{\frac{d_{1m}{n_1}^2 +  d_{2m}{n_2}^2}{\Lambda}}$). We assume $\Lambda = 400~nm$, $\delta = 10~nm$, $N=39$ and $m = 1,2,3,...(N-1)$ and subsequently, simulate the reflection spectrum using finite element technique (COMSOL Multiphysics). In the simulations, the periodic boundary condition is imposed in the transverse direction and a mapped mesh is used with a maximum element size $30~nm$. The material dispersion for $TiO_2$ and $SiO_2$ is obtained from \cite{Gao:15}. The blue-solid curve in Fig. \ref{Figure1}(b) represents the broadband ($\approx~80~THz$ broad) reflection spectrum for the CPC. The reflection spectrum is marked by discernible suppression of sharp reflectivity peaks outside the PBG that are distinct features of a finite periodic PCs. In fact, the transmission band (characterized by $\nu\leq~150~THz$ or $\nu\geq~260~THz$) is reasonably flat and transmission $T \geq~95\%$. A comparison of reflection spectrum (not shown here) for a periodic 1D-PC ($d_1 = d_2 = 200~nm$) shows that the PBG for CPC is approximately $\approx~40~THz$ broader than that for periodic PC. In order to appreciate this point, we represent the spectral variation in the real part of Bloch-wavevector ($K$) of periodic 1D-PCs with different values of $d_1$ such that $d_1+d_2=\Lambda = 400~nm$ for all the PCs in Fig. \ref{Figure1}(c). The flat region (for each $d_1$) depict the PBG (where $K$ is complex and its real part is constant $=\frac{2\pi}{\Lambda}$) for each PC. It is apparent that the PBGs for each PC (having different $d_1$ and $d_2$) approximately spans the wide spectral band from $150~THz$ to $250~THz$. This explains the origin of broad PBG as well as the suppression of reflectivity peaks outside the PBG in CPC. It is worthwhile to point out that the linear-chirp in PC ensures a symmetric variation of $\Delta k$ from a negative to a positive value and $\frac{d\Delta k}{dz} << {|\kappa_s|}^2$ at any $z$ (not shown here).
\begin{figure}[htbp]
\centering\includegraphics[width= \linewidth]{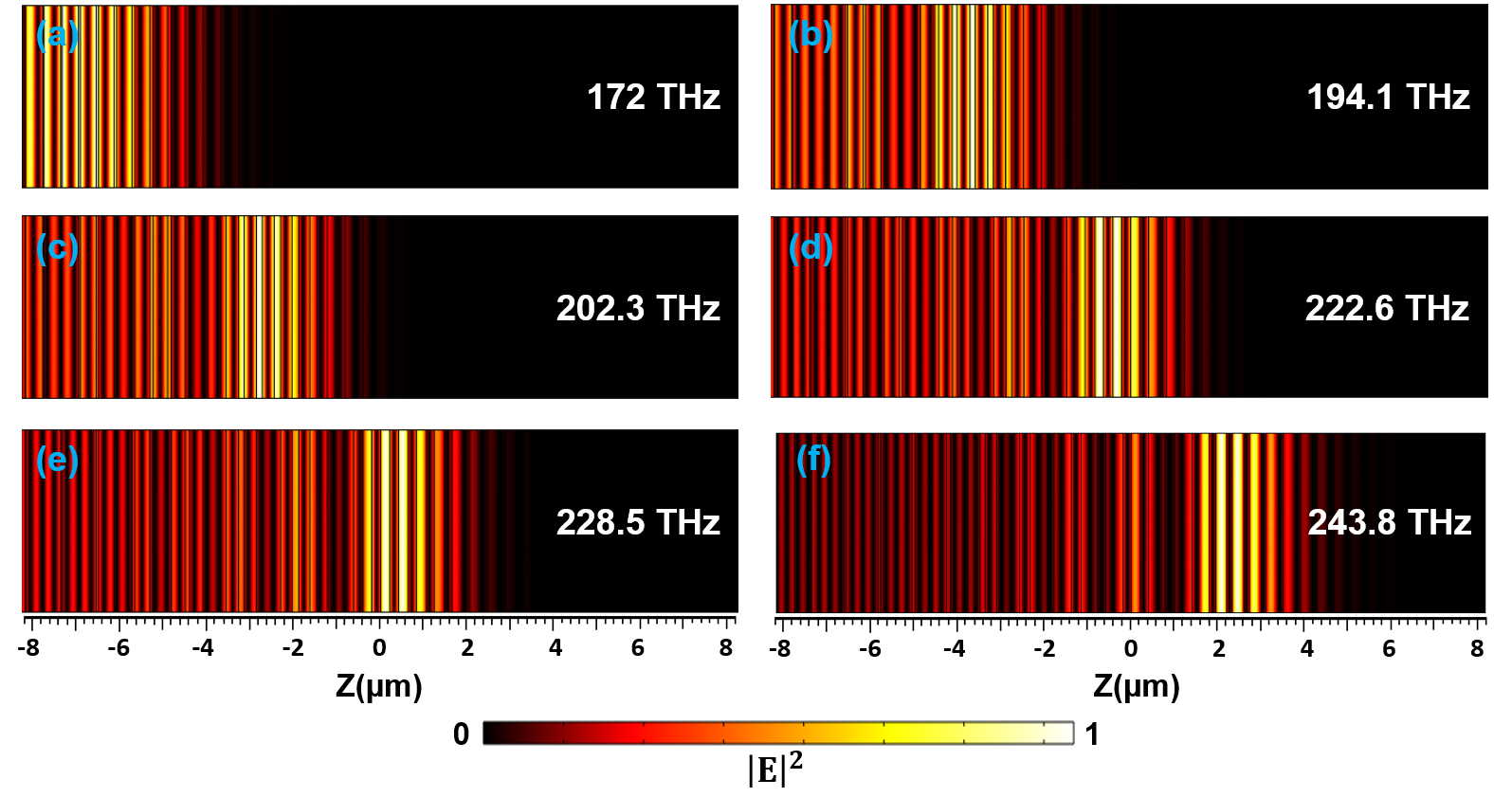}
\caption{Normalized mode-field intensity distribution ($\propto |E|^2$) for different OT modes (see Fig. \ref{Figure1}b) in Au-CPC geometry.}
\label{Figure2}
\end{figure}
In order to excite OT-like states/modes in the CPC, we assume a thin $Au$-layer ($d_m = 30~nm$) is placed adjacent to the first $TiO_2$ layer ($A$) as shown in Fig. \ref{Figure1}(a). The reflection spectrum is represented by the red-curve in Fig. \ref{Figure1}(b) which is characterized by sharp drop in reflectivity within the PBG of CPC. Such sharp resonances are a signature of OT modes in metallo-dielectric geometries and usually, the OT mode-field decays away from the metal-dielectric interface \cite{Shukla:18}. However, the mode-field distribution at a few OT resonant frequencies are shown in Fig. \ref{Figure2} which depict that the field localization at different resonant frequencies are spatially separated. The mode-field at smallest resonant frequency $\nu_{r1} = 172~THz$ resembles a conventional OT mode whereas the higher frequency modes (such as the ones at $\nu_{r2} = 194.1~THz$, $\nu_{r3} = 202.3~THz$, $\nu_{r4} = 222.6~THz$ etc.) are localized progressively away from the $Au$-$TiO_2$ interface (along $+z$ direction). In case of CPC, we obtain $11$ sharp OT resonances within the PBG. The origin of OT resonance is governed by the condition that $\phi_{PC}+\phi_{M}=2m\pi$ for $m=0,~1,~2,~3,...$ where $\phi_{PC}$ and $\phi_{M}$ are the phase acquired by the reflected beam from a semi-infinite PC and metal respectively \cite{PhysRevX.4.021017}. It is worth noting that $\phi_M$, in general, is negative for visible to mid-IR spectral band. However, the sign of $\phi_{PC}$ (in the PBG) exhibits a \emph{topological} connection \cite{PhysRevX.4.021017}. In fact, the sign of $\phi_{PC}$ (for a certain PBG) is dictated by the algebraic sum of topological (\emph{Zak}) phase for all the pass (transmission) bands below that PBG. Only those PBGs for which $\phi_{PC}$ is positive could support OT modes. In Fig. \ref{Figure1}(c), we observed that the CPC could be decomposed into multiple periodic PCs (with different $d_1$, $d_2$) and their overlapping PBGs result into a broader PBG for the CPC. Within the PBGs of these PCs, $\phi_{PC}$ traverses from [$0 \rightarrow \pi$] (anti-clockwise) or [$0\rightarrow-\pi$] (clockwise). From a topological perspective, PBGs in the former category only could support OT modes. Periodic PCs constituted using such $d_1$ and $d_2$ values support OT modes. Consequently, we obtain field localization due to OT mode formation in CPC near certain values of $d_{1}$, $d_{2}$ only. The field localization in Fig. \ref{Figure2} is in agreement with this argument and the OT mode-field amplitudes exhibit maxima at different $d_1$ and $d_2$.    
 
\begin{figure}[htbp]
\centering\includegraphics[width=\linewidth]{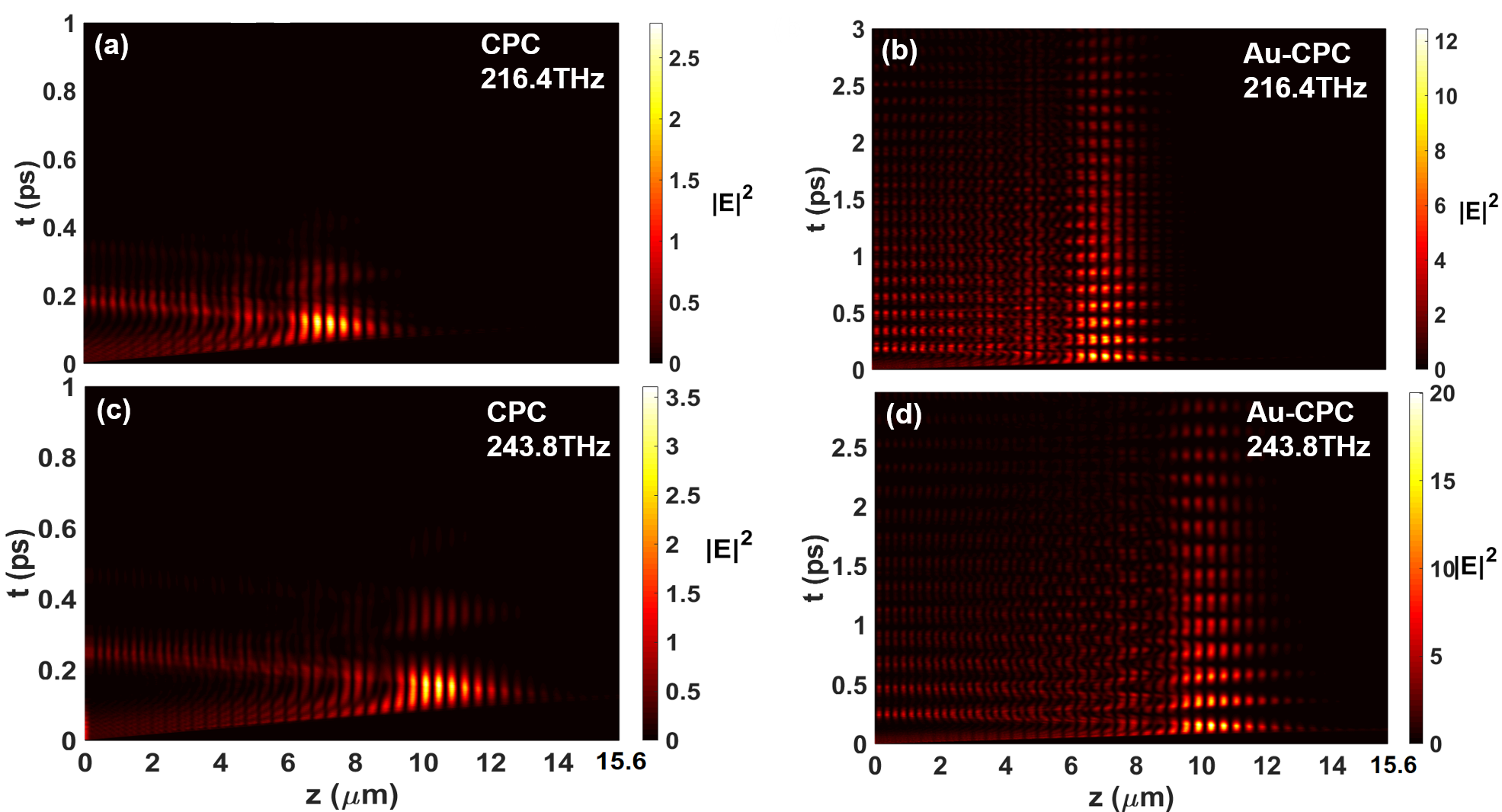}
\caption{Shows the simulated mode-field intensity ($\propto |E|^2$) distribution in the time-position plane along propagation ($z$) direction for a Gaussian pulse with width $100~fs$ centered at $\nu_r = 216.4~THz$ propagating through (a) CPC (b) Au-CPC geometry. Similar mode-field intensity distribution at $\nu_r=243.8~THz$ in (c) CPC and (b) Au-CPC.}
\label{Figure3}
\end{figure}
\begin{figure}[htbp]
\centering\includegraphics[width=\linewidth]{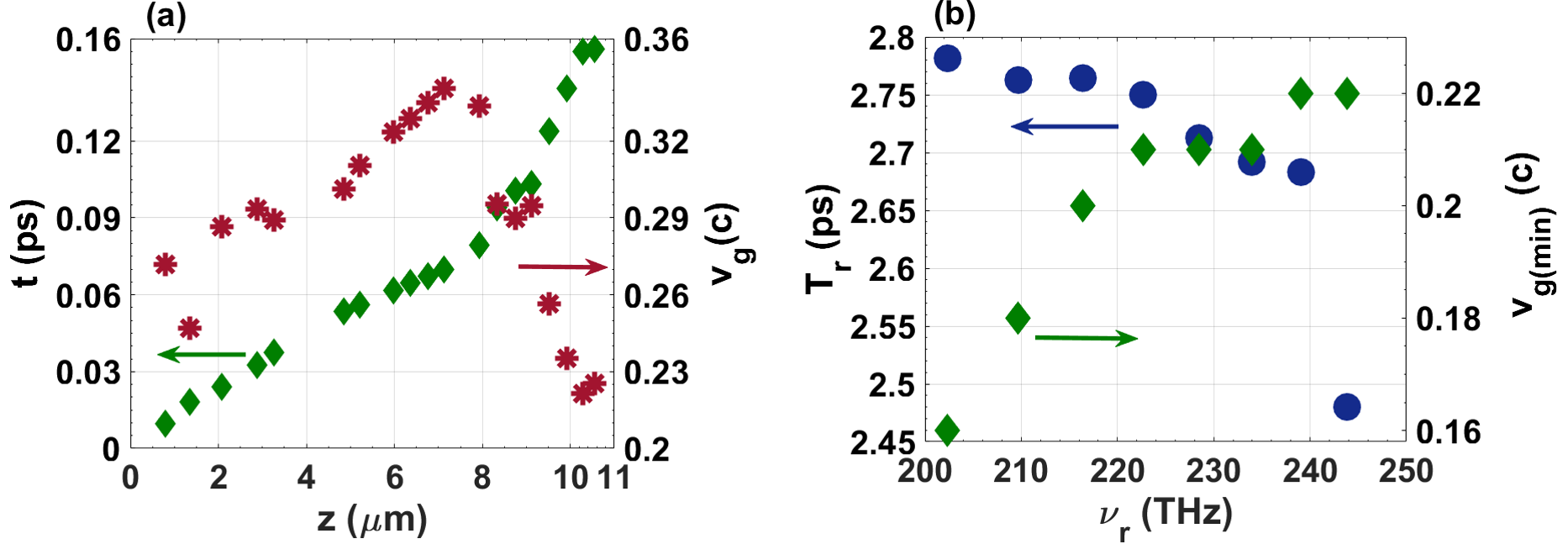}
\caption{a) shows the time of arrival (green square dots) at any $z$-coordinate in Au-CPC geometry for a $100~fs$ Gaussian pulse-peak and variation in $v_g$ (red asterisk-like dots) as a function of $z$ when the pulse central frequency is $\nu_r = 243.8~THz$. b) shows the trapped-mode lifetime (green square dots) and minimum group velocity $v_{g(min)}$ (blue circular dots) attained by different OT modes in Au-CPC.}
\label{Figure4}
\end{figure}

The $em$-field localization imply a drop in group velocity ($v_g$) of OT modes. In order to ascertain $v_g$, we allow, say a $100~fs$ Gaussian pulse centered at different OT resonant frequencies ($\nu_{r}$ in Fig. \ref{Figure1}(b)) to be incident on the thin $Au$-layer and perform a finite-difference time-domain (FDTD) simulation to obtain the evolution of the pulse. Figure \ref{Figure3} shows the result for such a simulation where a comparison is made between a bare CPC and CPC with Au-film (Au-CPC) at two different OT resonant frequencies (different $\nu_r$s). The time (in \emph{ps}) on the $y$-axis represents the time of arrival of pulse-peak at the respective locations. Figures \ref{Figure3}(a) and (c) show that the electric field-intensity of pulses at $\nu_{r} = 216.4~THz$ and $\nu_{r} = 243.8~THz$ in the bare CPC exhibit short-lived ($\approx~0.1-0.2~ps$) localization. On the other hand, identical pulses at the same OT resonant frequencies (in Au-CPC) exhibit localization over a prolonged period ($\geq~2~ps$) as shown in Figs. \ref{Figure3}(b) and (d). The time-domain simulations allow us to directly obtain $v_g$ for each OT mode by noting the time ($t$) taken by the pulse (peak) to arrive at any point ($z$) in the CPC (Fig. \ref{Figure4}(a)). By using this, the variation in $v_g$ for $\nu_r = 243.8~THz$ along the propagation direction is shown in Fig. \ref{Figure4}(a). From this figure, it could be ascertained that $v_g$ minimizes near the unit cell with $d_1 = 140~nm$ and $d_2 = 260~nm$ which is consistent with the field localization observed in Fig. \ref{Figure2} for the same OT resonant frequency. The minimum group-velocity ($v_{g(min)}$) attained by different OT modes (in Au-CPC) is shown in Fig. \ref{Figure4}(b). The $v_{g(min)}$ attained by OT mode at $\nu_{r} = 243.8~THz$ is $\approx~0.22c$. Figure \ref{Figure4}(b) shows that $v_{g(min)}$ varies over a range of $\approx~0.05c$ amongst all the OT resonant frequencies with the smallest value of $\approx~0.17c$ at $\nu_r = 202.3~THz$.  
The conventional rainbow trapping techniques utilize graded-indexed plasmonic gratings or waveguides which comprises ingrained absorptive metallic components. Although, they dictate the drop in $v_g$ through surface-plasmon excitation, they introduce unavoidable \emph{absorption} loss ($\alpha$) which puts a restriction on the time for which an optical pulse could be stored. Typically, the trapped-mode lifetime $T_r=\frac{1}{\alpha v_g}~\leq~1~ps$  for plasmon-based metallo-dielectric architectures. In our case, the Au-layer (plasmon-active metal film) at the terminating CPC layer leads to excitation of several OT modes. As shown in Fig. \ref{Figure2}, such modes are primarily localized within the CPC architecture and have a very small presence in the metal layer. Therefore, the dominant source of \emph{loss} in such OT modes (in Au-CPC) is the transmission loss due to finite length of photonic crystal. Since, different OT modes are spatially separated, they exhibit different transmission loss which could be estimated from the throughput modal power for a particular OT resonant frequency. Alternately, this also provides an estimate of the trapped-mode lifetime ($T_r$) for different OT modes. From the time-domain simulations in Figs. \ref{Figure3}(b) and (d), $T_r$ could be determined by monitoring the fall in pulse-peak to $1/e^2$-\emph{th} of its maximum value. For example, the OT mode at $nu_r = 243.8~THz$ has a trapped lifetime $T_r\approx 2.48~ps$. Due to different $v_{g(min)}$ for different OT modes, the trapped lifetime for OT modes varies and it is shown in Fig.\ref{Figure4}(b). $T_r$ maximizes to $2.78~ps$ at $\nu_r = 202.3~THz$ and varies by about $0.35~ps$ across the frequency tunable range. It is also apparent that the OT modes localized closer to the $Au$ layer exhibit longer trapped lifetimes which is essentially a consequence of smaller transmission losses.
\begin{figure}[htbp]
\centering\includegraphics[width=\linewidth]{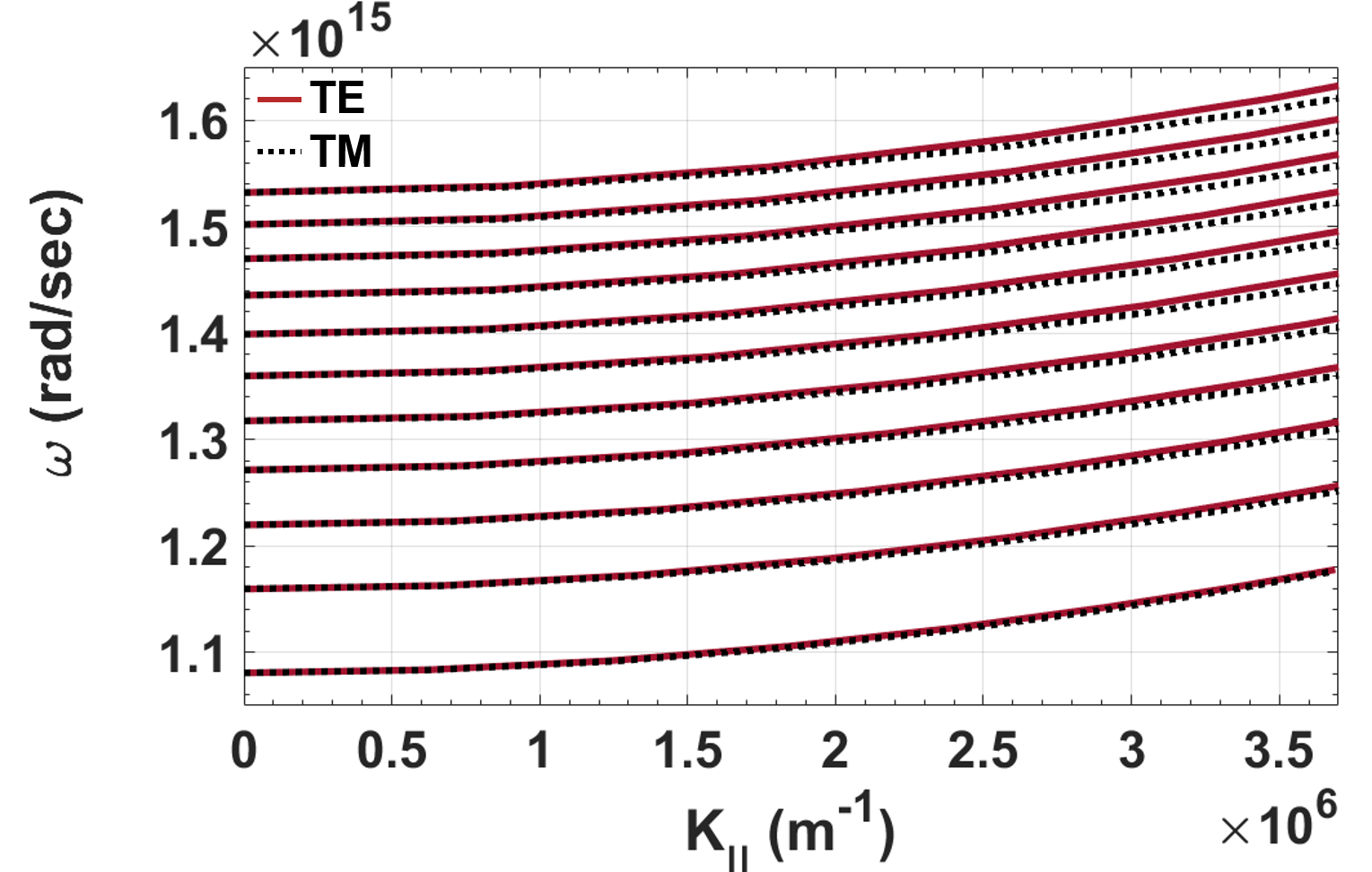}
\caption{Shows variation of angular frequency as a function of in-plane wavevector ($K_{\vert\vert}$) for the TE (solid red curve) and TM (dashed black curve) polarized OT modes in Au-CPC.}
\label{Figure5}
\end{figure} 

The dispersion characteristics of such resonant frequencies are shown in Fig. \ref{Figure5} where $K_{\vert\vert}$ represents the in-plane (or tangential) component of wavevector. All the OT modes exhibit a parabolic dispersion curve and the polarization degeneracy is lifted for higher values of in-plane component of the wavevector. This behaviour is similar to conventional OT modes localized at metal-dielectric interface. The low energy OT modes exhibit a small splitting between TE and TM polarizations at higher $K_{\vert\vert}$ values. The splitting tends to increase in case of higher energy OT modes. This is primarily due to discernible redistribution of the TE/TM polarized mode-fields for those OT modes which are spatially localized away (higher frequency) from the Au-$TiO_2$ interface.

In conclusion, we presented an optimally-designed CPC which closely follows the adiabatic constraints and consequently, leads to PBG broadening and multiple $\pi$-jumps in backscattered phase. This allows excitation of multiple sharp OT mode resonances in case of Au-CPC configuration. All the OT modes are spatially separated in the CPC and exhibit a topological connection. All such OT mode resonances provide a favorable platform for low-loss trapping of light with lifetimes as large as $\geq~2.8~ps$. In addition to this, the Au-CPC like architectures could be employed for designing plasmon-based sensing schemes. The possibility of manipulating the backscattered phase in such quasi-periodic all-dielectric PC allows to spatially structure optical beams and control their propagation characteristics. 

\bibliographystyle{unsrt}  
\bibliography{ref}  


\end{document}